\documentclass[journal]{IEEEtran}
\usepackage{cite}

\ifCLASSINFOpdf
  \usepackage[pdftex]{graphicx}
\else
\fi
\usepackage{amsmath}
\usepackage{array}

\usepackage{subfigure}

\usepackage[caption=false,font=footnotesize]{subfig}
\usepackage{dblfloatfix}

\usepackage{multirow}

\usepackage{algorithm}
\usepackage{algorithmic}  
\usepackage{color}

\newtheorem{remark}{Remark}

\begin{document}
%
\title{RHONN Modelling-enabled Nonlinear Predictive Control for Lateral Dynamics Stabilization of An In-wheel Motor Driven Vehicle}
%
%
%

\author{Hao~Chen, Junzhi~Zhang,
        and~Chen~Lv,~\IEEEmembership{Senior~Member,~IEEE} 

\thanks{Hao Chen and Chen Lv are with the School
of Mechanical and Aerospace Engineering, Nanyang Technological University, 639798,
Singapore, e-mails: chen.h@ntu.edu.sg, lyuchen@ntu.edu.sg}
\thanks{J. Zhang is with the School of Vehicle and Mobility, Tsinghua University, Beijing 100084, China. (E-mail: jzhzhang@mail.tsinghua.edu.cn)}

\thanks{Corresponding author: C. Lv}
}

%
%

\markboth{Journal of \LaTeX\ Class Files,~Vol.~14, No.~8, August~2015}%
{Shell \MakeLowercase{\textit{et al.}}: Bare Demo of IEEEtran.cls for IEEE Journals}
%



\maketitle

\begin{abstract}
Featuring the fast response and flexibility in control allocation, an electric vehicle with in-wheel motors is a good platform for implementing advanced vehicle dynamics control. Among many active safety functions of an in-wheel motor driven vehicle (IMDV), lateral stability control is a key technology, which can be realized through torque vectoring. To further advance the lateral stabilization performance of the IMDV, in this paper a novel data-driven nonlinear model predictive control (NMPC) is proposed based the recurrent high-order neural network (RHONN) modelling method. First, the new RHONN model is developed to represent vehicle’s nonlinear dynamic behaviors. Different from the conventional physics-based modelling method, the RHONN model forms high-order polynomials by neuron states to feature nonlinear dynamics. Based on the RHONN model, the steady-state responses of vehicle’s yaw rate and sideslip angle are iteratively optimized and set as the control objectives for low-level controller, aiming to improve the system robustness. Besides, a nonlinear model predictive controller is designed based on the RHONN, which is expected to improve the prediction accuracy during the receding horizon control. Further, a constrained optimization problem is formulated to derive the required yaw moment for vehicle lateral dynamics stabilization. Finally, the performance of the developed RHONN-based nonlinear MPC is validated on an IMDV in the CarSim/Simulink simulation environment. The validation results show that the developed approach outperforms the conventional method, and further improves the stable margin of the system. It is able to enhance the lateral stabilization performance of the IMDV under various driving scenarios, demonstrating the feasibility and effectiveness of the proposed approach. 
\end{abstract}


%

\section{Introduction}

\IEEEPARstart{A}{s} a typical type of electric vehicles, in-wheel motor driven vehicles (IMDV) have a distinctive powertrain configuration that cancels the transmission parts, including the clutch, torque converter, shafts and differential, and is directly driven by the hub motors. In addition, as each wheel can apply either a driving torque or a braking torque, the IMDV is able to realize torque vectoring (TV) through control allocation of the four wheels with following characteristics: 1) a higher transmission efficiency; 2) fast response and abundant feedback information of the electric powertrain; 3) redundant actuators with improved functional safety; 4) more flexible control ability. Particularly, as the motor torque on each wheel of an IMDV can be controlled independently, an additional yaw moment can be generated by torque vectoring without the intervention of the braking system, which guarantees the longitudinal driving demand simultaneously. All these properties make IMDVs an outstanding platform in terms of implementing vehicle dynamics control \cite{Ref1_kobayashi2017direct,Ref_ruiz2019front,Ref_siampis2013electric,Ref2_nguyen2019slip,Ref3_ghosh2015torque}.

Lateral stability control, realized by applying an additional yaw moment, is one of the key functions for vehicle active safety \cite{Ref4_tahouni2019novel,Ref5_oh2019yaw}. It can prevent the vehicle from excessive lateral acceleration or exceeding maximum friction force, assisting drivers to better control a vehicle and follow the desired trajectory under critical conditions. In an IMDV, the lateral stabilization, which can be easily realized by independently control of the motor torque, ensures both the handling maneuverability and stability, further improving the safety \cite{Ref6_sami2018improvement,Ref7_shibahata1993improvement,Ref_siampis2017real}. 

Many researchers have made great efforts in studying the lateral stability control for IMDVs in the past decades. Takao Kobayashi \emph{et al.} \cite{Ref8_kobayashi2018efficient,Ref9_kobayashi2017direct} established a 9 Degree-of-Freedom (9-DoF) dynamic model of an IMDV, and analyzed the relationship between the yaw moment and the power loss of the lateral slip. It was proved that the yaw moment control can effectively reduce the energy loss of an IMDV. The result showed that under steering conditions during vehicle acceleration, the minimum energy dissipation of the tires can be achieved only if the adhesion coefficient of each wheel reaches the same. Binh-Minh Nguyen \emph{et al.} \cite{Ref10_nguyen2016glocal} adopted a three-layer control method for vehicle torque vectoring. The upper-layer controller was used to derive the total driving force according to the expected speed and speed error. In the middle-layer, a LQR controller was designed to calculate the required total yaw moment based on the yaw rate and sideslip angle error. The lower controller minimized the total workload through a formulated constrained optimization problem to derive the torque command for each motor. Aria Noori Asiabar \emph{et al.} \cite{Ref11_asiabar2019direct} proposed a dual-layer control system, in which an adaptive sliding mode control was developed to yield the yaw moment. Ningyuan Guo \emph{et al.} \cite{Ref12_guo2020real,Ref13_guo2020fast} introduced a nonlinear model predictive control (NMPC) strategy to estimate the required additional yaw moment. Similar hierarchical control schemes with various control algorithms can also be found in \cite{Ref14_zhai2016electronic,Ref15_wong2016integrated}. The aforementioned works offer many feasible solutions to the lateral stability control of IMDVs, however, the control laws are highly dependent on the accuracy of the vehicle model. Besides, the linear tire models with massive empirical parameters are required for describing vehicle dynamics, but the linear assumption can easily fail in real world applications, and the accurate parameters of tire models can hardly be obtained.

Besides the controller design, the selection of control objectives is also very important. A reasonable control target is beneficial to the feedback control accuracy, which further enhances the control stability. The systems, reported in \cite{Ref16_ahmed2020vehicle,Ref17_hajiloo2021coupled}, adopted a two DoF linear single-track model to calculate the steady-state response of the yaw rate and vehicle sideslip angle, which were taken as the control objectives. Yonathan Weiss \emph{et al.} \cite{Ref18_weiss2018yaw} chose a first order filter to generate smooth and differentiable reference signals. Milad Jalali \emph{et al.} \cite{Ref19_jalali2017integrated} attempted to correct the desired yaw rate using vehicle sideslip angle, which was then set as the only reference signal. In this way, the lateral velocity can be dragged back to the safe region again when exceeding the stability boundary. Actually, the assumption of the linear single-track model is feasible in most of the normal operation conditions. However, when the lateral acceleration becomes large (i.e., the driver has a larger steering wheel angle input), the working point of the tire may enter the nonlinear region. Thus, the reference value derived from a linear tire model would not be consistent with vehicle’s nonlinear dynamic behavior. To overcome this issue, some investigations have been conducted. Alberto Parra \emph{et al.} \cite{Ref20_parra2020nonlinear} studied the energy-efficient understeering gradient under different lateral accelerations and longitudinal velocities, and further derived the reference table of yaw rate for control. The results suggested that under a same lateral acceleration, different adhesion coefficients would correspond to different optimal yaw rates. It is better to have the accurate information of road adhesion as prior knowledge for setting the reference yaw rate, while it is difficult to measure in real world applications. It can be found that, obtaining the steady-state yaw rate and vehicle sideslip angle without linear assumptions and too much prior knowledge is of great importance for improving the lateral stability control.

Thus, the implementation of lateral stability control for IMDVs via external yaw moment requires: 1) good robustness and high accuracy on vehicle dynamics modelling; 2) the precise real-time adaptation of the steady-state yaw rate and sideslip angle under varying conditions.

In recent years, the emerging data-driven modeling methods have attracted great attentions from both academia and industry. The learning-based methods no longer need detailed information of system physical structure and parameters, instead, it only requires data for mapping the relationship between the input and output of the system. Among many data-driven modeling methods, the recurrent high-order neural network (RHONN) is a promising one due to its unique features \cite{Ref21_djilali2019real,Ref22_ornelas2019neural}. Compared with standard recurrent neural network (RNN) which uses the linear combination of neuron states, the RHONN can better describe the nonlinear dynamics of a system by constructing high-order polynomials using neuron states and external inputs \cite{Ref23_sanchez2000adaptive}. In particular, it has been proven that if a sufficiently large number of the order is allowed for the polynomials of the RHONN model, it would be able to approximate any dynamical system to any degree of accuracy \cite{Ref24_kosmatopoulos1995high}.

As vehicle is a typical nonlinear system, it is worthwhile exploring new control methods by leveraging emerging learning-based tools to further improve vehicle dynamics performance \cite{Ref25_imani2018region,Ref26_dona2019stability}. To further advance the lateral stabilization performance of IMDVs, in this paper a novel RHONN modelling enabled nonlinear predictive control method is proposed. Based on the high-precision RHONN model, the vehicle lateral steady-state responses are obtained and set as references for the low-level controller. Besides, a RHONN-based nonlinear model predictive controller is designed for improving the lateral stability performance of the IMDV. Simulation validations of the proposed algorithms are then conducted on an IMDV under  various dynamic driving scenarios.

The reminder of this paper is organized as follows. Section \ref{sec2} illustrates the recurrent high-order neural network modelling approach. The design of the NMPC based on the RHONN for IMDV lateral stabilization is fully discussed in Section \ref{sec3}. Then, Section \ref{sec4} presents the simulation validation and results. Finally, conclusions are given in Section \ref{sec5}.

\section{The Recurrent High-Order Neural Network Modelling}\label{sec2}
RHONN is a type of feedforward neural network [26], in which the current outputs are associated to the previous states. The RHONN defines high-order polynomials through different neuron states and input data. It allows high-order interactions among neurons and inputs, and represents the nonlinear behaviors of a system accurately. The detailed modelling procedures of a RHONN are introduced as follows.
\subsection{The Structure of RHONN}
Considering a general nonlinear dynamical system, as shown in (\ref{eq1}):
\begin{equation}\label{eq1}
  \dot{\boldsymbol{x}}=\boldsymbol{f}(\boldsymbol{x}, \boldsymbol{u})
\end{equation}
where, $\boldsymbol{x} \in \mathbf{R}^{n}$ is $n$-dimensional states; $\boldsymbol{u} \in \mathbf{R}^{m}$ is $m$-dimensional inputs;$\boldsymbol{f}: \mathbf{R}^{n+m} \rightarrow \mathbf{R}^{m}$ is a smooth vector field of $\mathbf{C}^{\infty}$, indicating the state transitions. 

Design a RHONN to reconstruct the nonlinear dynamics of the system, which can be given by:
\begin{equation}\label{eq2}
  \dot{\chi}_{i}=-a_{i} \chi_{i}+b_{i}\left[\sum_{k=1}^{L} w_{i k} \prod_{j \in I_{k}} y_{j}^{d_{j}(k)}\right] \quad(i=1,2, \ldots, n)
\end{equation}
where, $\dot{\chi}_{i}$ is the state of the ${i}^{th}$ neuron, corresponding to system state expressed in (\ref{eq1}); $a_i$ and $b_i$ are coefficients, $\left\{I_{1}, I_{2}, \ldots, I_{L}\right\}$ is a collection of non-ordered subsets of $\{1,2, \ldots, n+m\}$, $w_{i k}$ represent the adjustable synaptic weight of the network, ${d_{j}(k)}$ are nonnegative integers. $y_j $ is either the state of a neuron passing through a hyperbolic tangent function or the external input, which is defined in the vector $\boldsymbol{y}$ as follows \cite{Ref24_kosmatopoulos1995high}.
\begin{equation}\label{eq3}
  \boldsymbol{y}=\left[\begin{array}{c}
  y_{1} \\
  \vdots \\
  y_{n} \\
  y_{n+1} \\
  \vdots \\
  y_{n+m}
  \end{array}\right]=\left[\begin{array}{c}
  S\left(x_{1}\right) \\
  \vdots \\
  S\left(x_{n}\right) \\
  u_{1} \\
  \vdots \\
  u_{m}
\end{array}\right]
\end{equation}
where, $\boldsymbol{u}=\left[u_{1}, u_{2}, \ldots u_{m}\right]^{\mathrm{T}}$ is the external input vector, which is the same as that in (\ref{eq1}). $S(\cdot)$ is a hyperbolic tangent function or a sigmoidal function. Here, the former is selected and given by:
\begin{equation}\label{eq4}
  S\left(x_{i}\right)=\mu_{i} \tanh \left(\beta_{i} x_{i}\right) \quad(i=1,2, \ldots, n)
\end{equation}
where, $\mu_{i}$ and $\beta_{i}$ are positive constants. As the signals sampled by the real system are all in discrete form, we adopt the discrete-time RHONN, as shown in (\ref{eq5}) \cite{Ref27_quintero2017neural,Ref28_quintero2018neural}:
\begin{equation}\label{eq5}
  \chi_{i, k+1}=\boldsymbol{W}_{i, k}^{T} \boldsymbol{\varphi}\left(\boldsymbol{\chi}_{k}, \boldsymbol{u}_{k}\right) \quad(i=1,2, \ldots, n)
\end{equation}
where, $\boldsymbol{\chi}_{k}=\left[\chi_{1, k}, \chi_{2, k}, \ldots \chi_{n, k}\right]^{\mathrm{T}}$ is the state vector of the neurons at time $k$, $\boldsymbol{u}_k=\left[u_{1,k}, u_{2,k}, \ldots u_{m,k}\right]^{\mathrm{T}}$ is the external input vector at time $k$, $\boldsymbol{W}_{i, k}=\left[w_{i 1, k}, w_{i 2, k}, \ldots w_{i L, k}\right]^{\mathrm{T}}$is the respective on-line adapted weight vector of ${i}^{th}$ neuron at time $k$. $\boldsymbol{\varphi}: \mathbf{R}^{n+m} \rightarrow \mathbf{R}^{L}$ is the a smooth vector field of $\mathbf{C}^{\infty}$ with each element formed by a high-order polynomial, which can be expressed as:
\begin{equation}\label{eq6}
  \boldsymbol{\varphi}\left(\boldsymbol{\chi}_{k}, \boldsymbol{u}_{k}\right)=\left[\begin{array}{c}
  \varphi_{1, k} \\
  \varphi_{2, k} \\
  \vdots \\
  \varphi_{L, k}
  \end{array}\right]=\left[\begin{array}{c}
  \Pi_{j \in I_{1}} \xi_{j, k}^{d_{j}(1)} \\
  \Pi_{j \epsilon I_{2}} \xi_{j, k}^{d_{j}(2)} \\
  \vdots \\
  \Pi_{j \epsilon I_{L}} \xi_{j, k}^{d_{j}(L)}
  \end{array}\right]
\end{equation}
where, $\xi_{j, k}$ is similar to the $y_j$ given by (\ref{eq3}) and defined in $\boldsymbol{\xi}$ as follows:
\begin{equation}\label{eq7}
  \boldsymbol{\xi}\left(\boldsymbol{\chi}_{k}, \boldsymbol{u}_{\boldsymbol{k}}\right)=\left[\begin{array}{c}
  \xi_{1, k} \\
  \vdots \\
  \xi_{n, k} \\
  \xi_{n+1, k} \\
  \vdots \\
  \xi_{n+m, k}
  \end{array}\right]=\left[\begin{array}{c}
  S\left(\chi_{1, k}\right) \\
  \vdots \\
  S\left(\chi_{n, k}\right) \\
  u_{1, k} \\
  \vdots \\
  u_{m, k}
  \end{array}\right]
\end{equation}

Especially, if partial information of the system dynamics is known, then (\ref{eq5}) can be re-written as:
\begin{equation}\label{eq8}
 \begin{array}{c}
 \chi_{i, k+1}=\boldsymbol{W}_{i, k}^{\prime T} \boldsymbol{\psi}\left(\boldsymbol{\chi}_{k}, \boldsymbol{u}_{k}\right)+\boldsymbol{W}_{i, k}^{T} \boldsymbol{\varphi}\left(\boldsymbol{\chi}_{k}, \boldsymbol{u}_{k}\right)\\
 (i=1,2, \ldots, n)
 \end{array}
\end{equation}
where, $\boldsymbol{W}_{i, k}^{\prime}=\left[w_{i 1, k}^{\prime}, w_{i 2, k}^{\prime}, \ldots w_{i s, k}^{\prime}\right]^{\mathrm{T}}$ is the fixed weight vector of the ${i}^{th}$ neuron at time $k$, $\boldsymbol{\psi}: \mathbf{R}^{n+m} \rightarrow \mathbf{R}^{S}$ is the a smooth vector field of $\mathbf{C}^{\infty}$, which is identical to $\boldsymbol{\varphi}$. The high-order polynomial in $\boldsymbol{\psi}$ reflects the system dynamics, as shown in (\ref{eq9}).
\begin{equation}\label{eq9}
  \boldsymbol{\psi}\left(\boldsymbol{\chi}_{k}, \boldsymbol{u}_{k}\right)=\left[\begin{array}{c}
  \psi_{1, k} \\
  \psi_{2, k} \\
  \vdots \\
  \psi_{s, k}
  \end{array}\right]=\left[\begin{array}{c}
  \Pi_{j \epsilon I_{l 1}} \xi_{j, k}^{d_{j}(l 1)} \\
  \Pi_{j \epsilon I_{l 2}} \xi_{j, k}^{d_{j}(l 2)} \\
  \vdots \\
  \Pi_{j \epsilon I_{l s}} \xi_{j, k}^{d_{j}(l s)}
  \end{array}\right]
\end{equation}

\begin{figure}[!t]
\centering
\includegraphics[width=2.5in]{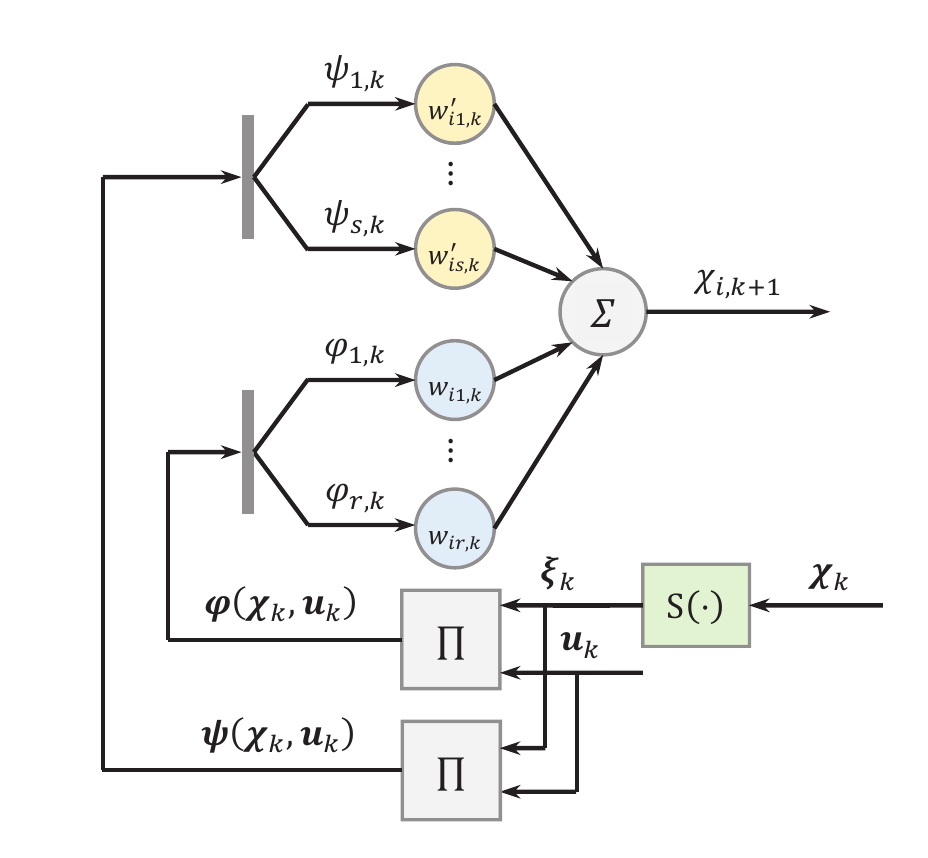}
\caption{The architecture of an ${i}^{th}$ order discrete-time RHONN model.}
\label{fig1}
\end{figure}

The definitions of $\boldsymbol{W}_{i, k} \in \mathbf{R}^{r}$ and $\boldsymbol{\varphi}: \mathbf{R}^{n+m} \rightarrow \mathbf{R}^{r}$ shown in (\ref{eq9}) are the same as that illustrated in (\ref{eq6}), but the dimension of $\boldsymbol{\varphi}$ is reduced, i.e., $r \leq L$. The structure of an ${i}^{th}$ order discrete-time RHONN is depicted in Fig. \ref{fig1}.

\begin{remark}
In terms of the architecture design of the RHONN, the fixed part, i.e., $\boldsymbol{\psi}$, consists of these high-order polynomials with known parameters. The remaining part, i.e., $\boldsymbol{\varphi}$, features the unknown part of the system, including model uncertainties and disturbances. In this way, the prior knowledge of system dynamics is fully utilized. Therefore, the size of the RHONN can be reduced without serious computational concerns.
\end{remark}

\subsection{Weight Optimization via Extended Kalman Filter-based Learning}\label{sec2.2}

As the weight matrices influence the neuron states and performance of the RHONN, it is essential to identify and update the weights. There are several online adaptive learning algorithms for the weights update: the filter regressor RHONN \cite{Ref29_rovithakis2012adaptive}, filter error RHONN \cite{Ref29_rovithakis2012adaptive}, robust learning algorithm \cite{Ref29_rovithakis2012adaptive}, Extended Kalman Filter (EKF)-based learning algorithm \cite{Ref30_alanis2007discrete}, and et.al. Among these approaches, the EKF-based learning is promising due to its high computation efficiency and fast convergence. Thus, we adopt it for the weight parameter online identification in this work.

Define the error of the  ${i}^{th}$ state $e_{i, k}$, as shown in (\ref{eq10}).
\begin{equation}\label{eq10}
  e_{i, k}=x_{i, k}-\chi_{i, k} \quad(i=1,2, \ldots, n)
\end{equation}
where, $x_{i, k}$ is obtained from the physical plant. It can be either measured by sensors or estimated by using state observer. $\chi_{i, k}$ is the state of the ${i}^{th}$ neuron.
\begin{figure*}[!t]
  \centering
  \includegraphics[width=5.3in]{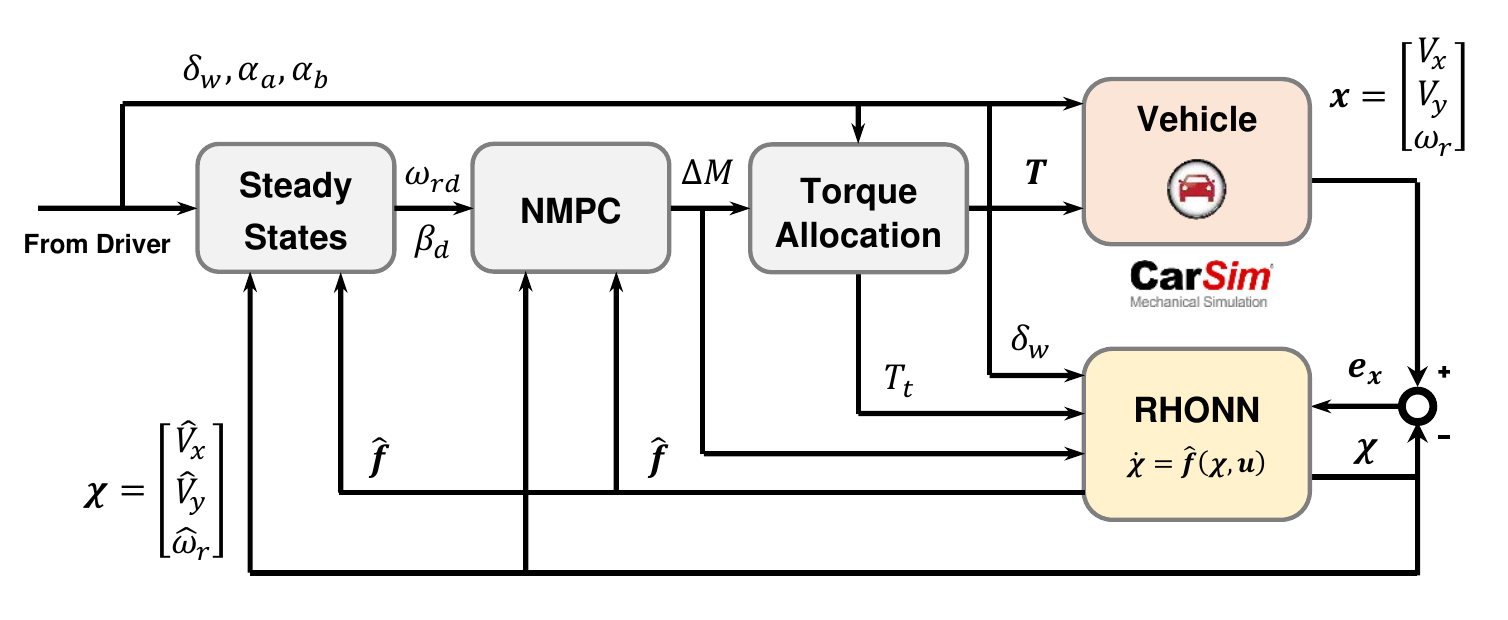}
  \caption{The block diagram of the RHONN-based lateral stability controller for IMDVs.}
  \label{fig2}
\end{figure*}

The weight matrix of the ${i}^{th}$ neuron at time $k$ can be updated using the following EKF-based learning mechanism:
\begin{small}
\begin{equation}\label{eq11}
  \begin{array}{c}
  \boldsymbol{W}_{i, k+1} =\boldsymbol{W}_{i, k}+\zeta_{i} \boldsymbol{K}_{i, k} e_{i, k} \\
  \boldsymbol{K}_{i, k} =\boldsymbol{P}_{i, k} \boldsymbol{H}_{i, k}\left[\boldsymbol{R}_{i, k}+\boldsymbol{H}_{i, k}^{T} \boldsymbol{P}_{i, k} \boldsymbol{H}_{i, k}\right]^{-1}\\
  \boldsymbol{P}_{i, k+1} =\boldsymbol{P}_{i, k}-\boldsymbol{K}_{i, k} \boldsymbol{H}_{i, k}^{\mathrm{T}} \boldsymbol{P}_{i, k}+\boldsymbol{Q}_{i, k}\\
  (i=1,2, \ldots, n) 
  \end{array}
\end{equation}
\end{small}
where, $\zeta_{i}$ is a learning parameter, $\boldsymbol{R}_{i, k} \in \mathbf{R}^{1}$ is covariance matrix associated with measurement $x_{i, k}$; $\boldsymbol{Q}_{i, k} \in \mathbf{R}^{r \times r}$ is covariance matrix associated with the weight vector $\boldsymbol{W}_{i,k}$. And $\boldsymbol{H}_{i,k}$ is the Jacobian matrix, which can be given by
\begin{equation}\label{eq12}
  \boldsymbol{H}_{i, k}=\left[\frac{\partial \chi_{i, k}}{\partial w_{i, k}}\right]^{T}=\left[\varphi_{i, k}\right]^{T} \in \mathbf{R}^{r} \quad(i=1,2, \ldots, n)
\end{equation}

Noted that, different from learning algorithms in other neural networks, the EKF-based learning method in (\ref{eq11}) adjusts $\boldsymbol{W}_{i,{k+1}}$ at each step. This also enables the RHONN model to quickly construct the nonlinear dynamics of the system. 

\section{RHONN-based Nonlinear Predictive Controller Design for Lateral Stabilization}\label{sec3}
\subsection{The high-level structure of the RHONN-based NMPC}
The lateral stabilization enhances the vehicle's lateral stability performance by external yaw moment while maintaining the driver's intentions. Because the yaw rate and sideslip angle are two indicators that represent lateral dynamics of vehicles, they are selected here as the control objectives. The lateral stability control scheme based on RHONN developed for IMDVs is shown in Fig. \ref{fig2}.
  
Here, a general RHONN model for planar dynamics of the IMDV can be expressed as:
\begin{equation}\label{eq13}
  \dot{\boldsymbol{\chi}}=\hat{\boldsymbol{f}}(\boldsymbol{\chi}, \boldsymbol{u}) 
\end{equation}
where, $\hat{\boldsymbol{f}}: \mathbf{R}^{n+m} \rightarrow \mathbf{R}^{m}$ is a RHONN mapping function. $\boldsymbol{\chi}$ is the state vector, and $\boldsymbol{u}$ is the system input, as shown in the following (\ref{eq14}) .
\begin{equation}\label{eq14}
  \begin{array}{c}
  \boldsymbol{\chi}=\left[\hat{V}_{x}, \hat{V}_{y}, \hat{\omega}_{r}\right]^{\mathrm{T}}  \\
  \boldsymbol{u}=\left[T_t,\Delta M, \delta_{w}\right]^{\mathrm{T}}
  \end{array}
\end{equation}
where, $\hat{V}_{x}$, $\hat{V}_{y}$ and $\hat{\omega}_{r}$ are the estimates of the longitudinal velocity, lateral velocity and yaw rate by the RHONN model, respectively. In Fig. \ref{fig2}, $\alpha_{a}$ is the accelerator signal, $\alpha_{b}$ is the braking signal, and $\delta_{w}$ is the steering wheel angle input, from driver's maneuver. $T_t$ is the total driving torque, $\Delta M$ is the external yaw moment, which refers to the torque difference between left and right wheels, and $\boldsymbol{T}=\left[T_{1}, T_{2}, T_{3}, T_{4},\right]^{\mathrm{T}}$ is the torque vector, $i=1,2,3,4$  represent the torque command of front-left, front-right, rear-left and rear-right wheels, respectively.

The RHONN model reconstructs the vehicle dynamic characteristics in a data-driven manner. The steady-state values of yaw rate and vehicle sideslip angle, calculated based on the RHONN model, are regarded as the control objectives of the downstream controller. Then, the NMPC, which takes the RHONN as the predictive model, derives the external yaw moment for lateral stabilization. Further, the torque allocation module distributes the total driving torque and the external yaw moment to each wheel. The calculation of the total driving torque and allocation algorithm have been reported in \cite{Ref31_chen2020dynamic}.

\begin{figure}[!b]
  \centering
  \includegraphics[width=2.5in]{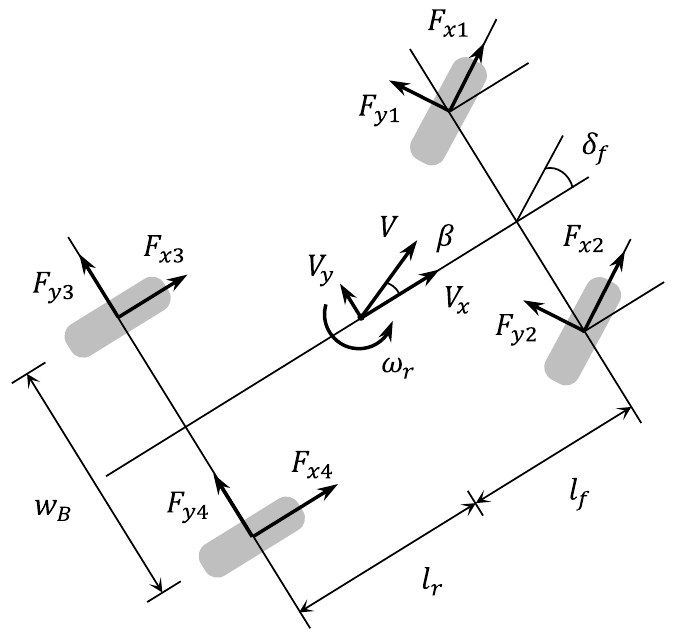}
  \caption{Four-wheel vehicle planar dynamics model of the IMDV.}
  \label{fig3}
\end{figure}

\subsection{The RHONN model for vehicle planar dynamics}
The planar dynamics model is a simple and practical model, which can be used to determine or predict vehicle behaviors. A typical four-wheel vehicle planar dynamics model of the IMDV is shown in Fig. \ref{fig3}.

With an assumption of the small steering angle, the Newton–Euler equations of the planar motion are:
\begin{equation}\label{eq15}
  \left\{\begin{aligned}
  \dot{V}_{x} &= \frac{1}{m r}\left(T_{1}+T_{2}+T_{3}+T_{4}\right) -\frac{I_{w}}{m r}\left(\dot{w}_{1}+\dot{w}_{2}+\dot{w}_{3}+\dot{w}_{4}\right)\\
  &-\frac{1}{m}\left(F_{y 1}+F_{y 2}\right) \delta_{f}+V_{y} \omega_{r} \\
  \dot{V}_{y} &=\frac{1}{m r}\left(T_{1}+T_{2}\right) -\frac{I_{w}}{m r}\left(\dot{w}_{1}+\dot{w}_{2}\right)\delta_{f} \\
  &+\frac{1}{m}\left(F_{y 1}+F_{y 2}+F_{y 3}+F_{y 4}\right)-V_{x} \omega_{r} \\
  \dot{\omega}_{r} &=\frac{w_{B}}{2 I_{z}}\frac{1}{r}\left(T_{2}+T_{4}-T_{1}-T_{3}\right) \\
  &-\frac{w_{B}}{2 I_{z}}\frac{I_{w}}{r}\left(\dot{w}_{2}+\dot{w}_{4}-\dot{w}_{1}-\dot{w}_{3}\right)\\
  &+\left(F_{y 1}+F_{y 2}\right) \frac{l_{f}}{I_{z}}-\left(F_{y 3}+F_{y 4}\right)\frac{l_{r}}{I_{z}}
  \end{aligned}\right.
\end{equation}
where, $F_{y i}(i=1,2,3,4)$ is the lateral tire force,
$\dot{w}_{i}(i=1,2,3,4)$ is the rotational acceleration, $\delta_{f}$ is the steering angle of the front wheels, $m$ is the vehicle mass, $I_{w}$ is the wheel mass moment, $I_{z}$ is the mass moment, $r$ is the rolling radius, $l_{f}$ is the distance from the front axle to c.g., $l_{r}$ is the distance from the rear axle to c.g., and $w_{B}$ is the track width. 

According to the definition of $T_t$ and $\Delta M$, we have:
\begin{equation}\label{eq16}
  \left\{\begin{aligned}
    T_{t}(k) &=T_{1}(k)+T_{2}(k)+T_{3}(k)+T_{4}(k) \\
  \Delta M(k)&=T_{2}(k)+T_{4}(k)-T_{1}(k)-T_{3}(k)
  \end{aligned}\right.
\end{equation}

$T_t$ is interpreted by the accelerator and braking signals from the driver. Besides, $\Delta M$ does not correspond to the actual yaw moment acting on the body, but it can indirectly correct the lateral dynamics and act as a control variable. 

Similar to (\ref{eq8}), the data-driven model based on RHONN for vehicle planar dynamics can be expressed by:
\begin{equation}\label{eq17}
  \left\{\begin{array}{c}
  \hat{V}_{x}(k+1)=\frac{1}{m r} T_{t}(k)+\boldsymbol{W}_{x, k}^{T} \boldsymbol{\varphi}\left(\boldsymbol{\chi}_{k}, \boldsymbol{u}_{k}\right) \\
  \hat{V}_{y}(k+1)=\boldsymbol{W}_{y, k}^{T} \boldsymbol{\varphi}\left(\boldsymbol{\chi}_{k}, \boldsymbol{u}_{k}\right) \\
  \hat{\omega}_{r}(k+1)=\frac{w_{B}}{2 I_{z}r} \Delta M(k)+\boldsymbol{W}_{r, k}^{T} \boldsymbol{\varphi}\left(\boldsymbol{\chi}_{k}, \boldsymbol{u}_{k}\right)
  \end{array}\right.
\end{equation}
 where, $\boldsymbol{W}_{x, k} \in \mathbf{R}^{15}, \boldsymbol{W}_{y, k} \in \mathbf{R}^{15}$, and $\boldsymbol{W}_{r, k} \in \mathbf{R}^{15}$ are the weight matrices for the longitudinal velocity, lateral velocity, and yaw rate, respectively. $\boldsymbol{\varphi}\left(\boldsymbol{\chi}_{k}, \boldsymbol{u}_{k}\right)$ is the high-order polynomial vector, which is illustrated in details in \emph{Appendix}.

\begin{remark}
It should be noted that (\ref{eq17}) is not strictly equivalent to (\ref{eq15}). In (\ref{eq15}), $\delta_{f}$ is assumed with a small value, whereas (\ref{eq17}) has no constraint on this criterion. In addition, the learning part includes not only the external disturbances but also the model uncertainties. All these features make the RHONN model advantageous over the conventional physics-based modelling methods.
\end{remark}
 
\subsection{RHONN-based lateral steady-state responses}\label{sec3.3}
The steady-state lateral responses of a vehicle include the unchanged lateral velocity and yaw rate, which are caused by a step input of the steering wheel angle under a constant longitudinal velocity. They are further set as the tracking objectives for the vehicle lateral stabilization. The discretized form of the steady-state lateral responses is expressed as:
\begin{equation}\label{eq18}
  \left\{\begin{array}{l}
  \hat{V}_{y d}(k+1)=\hat{V}_{y d}(k) \\
  \hat{\omega}_{r d}(k+1)=\hat{\omega}_{r d}(k)
  \end{array}\right.
\end{equation}
here, $\hat{V}_{yd}$ and $\hat{\omega}_{rd}$ are the steady lateral velocity and yaw rate, respectively.

As mentioned before, many scholars use the linear time-invariant model to generate the steady-state responses as the tracking targets, but the vehicle dynamic behaviors are highly nonlinear and time-varying. The RHONN model, $\hat{\boldsymbol{f}}(k)$), can approximate the nonlinear system behaviors in real-time. As such, the steady lateral velocity and yaw rate at time $k$ can be derived by the known $\hat{\boldsymbol{f}}(k)$), which further improves the accuracy of steady-state responses. Therefore, with consideration of the physical limits of the lateral velocity and yaw rate, the RHONN-based lateral vehicle steady-state responses can be formulated as a constrained nonlinear optimization problem from (\ref{eq17}) and (\ref{eq18}): 
\begin{equation}\label{eq19}
  \begin{array}{c}
  \min\limits_{\hat{V}_{y d}(k)\atop\hat{\omega}_{r d}(k)} \parallel\hat{V}_{y d}(k+1)-\hat{V}_{y d}(k)\parallel+\eta\parallel\hat{\omega}_{r d}(k+1)-\hat{\omega}_{r d}(k)\parallel \\
  \text { s.t. }\left\{\begin{array}{c}
  \hat{V}_{y d}(k+1)=\boldsymbol{W}_{y, k}^{T} \boldsymbol{\varphi}\left(\boldsymbol{\chi}_{k}, \boldsymbol{u}_{k}\right) \\
  \hat{\omega}_{r d}(k+1)=\boldsymbol{W}_{r, k}^{T} \boldsymbol{\varphi}\left(\boldsymbol{\chi}_{k}, \boldsymbol{u}_{k}\right) \\
  \hat{V}_{ymin}(k) \leq \hat{V}_{yd}(k) \leq \hat{V}_{ymax}(k) \\
  \hat{\omega}_{rmin}(k) \leq \hat{\omega}_{rd}(k) \leq \hat{\omega}_{rmax}(k)
  \end{array}\right.
  \end{array}
\end{equation}
where, $\eta$ is a tuning parameters for regulating the magnitude difference between the longitudinal velocity and yaw rate; $\hat{V}_{ymin}(k)$ and $\hat{V}_{ymax}(k)$ are the upper and lower limits of the lateral velocity at time $k$, respectively. $\hat{\omega}_{rmin}(k)$ and $\hat{\omega}_{rmax}(k)$ are the upper and lower limit of the yaw rate at time $k$, respectively. 

\begin{algorithm}[htb]
  \caption{ The "neighbors-searching" method for solving the equilibrium point.}
  \label{alg:Solution}
  \begin{algorithmic}[1] 
  \REQUIRE ~~\\ 
      The lateral velocity at previous step, $V_{y}(k-1)$;\\
      The yaw rate at previous step, $\omega_{r}(k-1)$;\\
      The initial searching radius, $\boldsymbol{r}_{s0}$;\\
      The initial searching area, $\boldsymbol{N}_{\varepsilon0}=\boldsymbol{0}$;\\
      The threshold value, $\mathcal{E}_{t}$
  \ENSURE ~~\\ 
      The reference lateral velocity at current step, $\hat{V}_{yd}(k)$;\\
      The reference yaw rate at current step, $\hat{\omega}_{rd}(k)$;\\
      \textbf{\emph{Initialization}} $\boldsymbol{C}_{0}=[V_{y}(k-1),\omega_{r}(k-1)]^{\mathrm{T}}$, $\boldsymbol{r}_{s}=\boldsymbol{r}_{s0}$, $\boldsymbol{N}_{\varepsilon p}=\boldsymbol{N}_{\varepsilon0}$
      \STATE Setting the neighbor area $\boldsymbol{N}_{\varepsilon}\left(\boldsymbol{C}_{0}, \boldsymbol{r}_{s}\right)$ with the center $\boldsymbol{C}_{0}$ and the radius $\boldsymbol{r}_{s}$ ;
      \STATE Conducting discretization on $\boldsymbol{N}_{\varepsilon}\left(\boldsymbol{C}_{0}, \boldsymbol{r}_{s}\right)$ -- $\boldsymbol{N}_{\varepsilon p}$ with a pre-defined interval ;
      \STATE Obtaining the feasible set $\boldsymbol{F}_{S}$ and its corresponding costs set $\boldsymbol{V}_{s}$;
      \STATE Choosing the optimal point $F_{s}^{*}=\left[V_{y}^{*}, \omega_{r}^{*}\right]^{\mathrm{T}}$ and its cost $V_{S}^{*}$;
      \STATE \textbf{If} $V_{S}^{*} \leq \mathcal{E}_{t}$ \textbf{then} $\hat{V}_{yd}(k)=V_{y}^{*}, \hat{\omega}_{rd}(k)=\omega_{r}^{*}$;\\
             \textbf{Else do} $\boldsymbol{N}_{\varepsilon p}=\boldsymbol{N}_{\varepsilon}, \boldsymbol{r}_{s}=\boldsymbol{r}_{s}+\Delta \boldsymbol{r}_{s}$ \textbf{and} back to step 2;
  \RETURN $\hat{V}_{yd}(k),\hat{\omega}_{rd}(k)$; 
  \end{algorithmic}
\end{algorithm}

The "neighbors-searching" method is introduced to solve (\ref{eq19}) in Algorithm \ref{alg:Solution}, which finds the solution close to the previous states and avoid multiple equilibriums problem. The maximum searching radius is constrained by the limits shown in (\ref{eq19}). Here are two metrics for Algorithm \ref{alg:Solution} : 1) the threshold value, $\mathcal{E}_{t}$, which ensures the solution existence for (\ref{eq19}); 2) the algorithm identifies the nearby equilibrium pointing to previous states as the references, reducing the difficulty for tracking tasks and preventing aggressive control actions. 

According to the solution of $\hat{\omega}_{rd}(k)$ and $\hat{V}_{yd}(k)$, the desired vehicle sideslip angle is defined as:
\begin{equation}\label{eq20}
  \hat{\beta}_{d}(k)=\frac{\hat{V}_{y d}(k)}{\hat{V}_{x}(k)}
  \end{equation}

As discussed above, the RHONN-based steady yaw rate and vehicle sideslip angle are more reasonable than conventional methods, which are further set as the control objectives for the lower controller.

\subsection{Nonlinear model predictive control law}
Predictive control has the characteristics of "feedforward + feedback" promising for dynamic target tracking problems. The main objective of the NMPC for lateral stabilization is to generate an external yaw moment to follow the desired yaw rate and sideslip angle, as defined in section \ref{sec3.3}. The predictive model fidelity would affect the control performance of the NMPC controller. Since the RHONN model can accurately approximate the system dynamics, it is taken as the predictive model for the NMPC controller. 

The RHONN-based predictive model is:
\begin{equation}\label{eq21}
  \begin{aligned}
  \boldsymbol{\chi}(k+i \mid k)&=\left[\begin{array}{c}
  \hat{V}_{x}(k+i \mid k) \\
  \hat{V}_{y}(k+i \mid k) \\
  \hat{\omega}_{r}(k+i \mid k)
  \end{array}\right] \\
  &=\left[\begin{array}{c}
  \frac{1}{m r} T_{t}(k)+\boldsymbol{W}_{x, k}^{T} \boldsymbol{\varphi}(k+i-1 \mid k) \\
  \boldsymbol{W}_{y, k}^{T} \boldsymbol{\varphi}(k+i-1 \mid k) \\
  \boldsymbol{W}_{r, k}^{T} \boldsymbol{\varphi}(k+i-1 \mid k)
  \end{array}\right]\\
  &+\left[\begin{array}{c}
  0 \\
  0 \\
  \frac{w_{B}}{2 I_{z}r}
  \end{array}\right] \Delta M(k+i-1 \mid k) \quad(i=1,2,3)
  \end{aligned}
\end{equation}
where, $\boldsymbol{\varphi}(k+i-1 \mid k)=\boldsymbol{\varphi}(\boldsymbol{\chi}(k+i-1 \mid k), \delta_{w}(k+i-1 \mid k))(i=1,2,3)$, and $\boldsymbol{\chi}(k+i \mid k)$ is the predictive state vector at $i$ steps ahead from current time $k$. Besides, the weights and the total driving torque remain unchanged in the prediction horizon.

The cost functions are selected as a quadratic form for tracking the errors of yaw rate and vehicle sideslip angle, respectively. The control and prediction horizon are set as 3 steps to coordinate with the time delay of the discrete-time system in (\ref{eq22}).
\begin{equation}\label{eq22}
  \begin{aligned}
  J_{1}&=\sum_{i=1}^{3} q_{i}\left[\hat{\omega}_{r d}(k+i)-\hat{\omega}_{r}(k+i \mid k)\right]^{2} \\
  J_{2}&=\sum_{i=1}^{3} r_{i}\left[\hat{\beta}_{d}(k+i)-\hat{\beta}(k+i \mid k)\right]^{2}
  \end{aligned}
\end{equation}
where, $q_{i}$ and $r_{i}(i=1,2,3)$ are weight coefficients for each prediction horizon.
Here, we set $\hat{\omega}_{r d}(k+i) = \hat{\omega}_{r d}(k)$, $\hat{\beta}_{r d}(k+i) = \hat{\beta}_{r d}(k)$, and $q_{i}= 100$, $r_{i}=1000$  for $i=1,2,3 $.

Besides, the external yaw moment should be constrained by the motor capacity:
\begin{equation}\label{eq23}
  \Delta M_{\min} \leq \Delta M(k+i-1 \mid k) \leq \Delta M_{\max } \quad(i=1,2,3)
\end{equation}
where, $\Delta M_{\max }$ and $\Delta M_{\min}$ are the upper limit and lower limit of the external yaw moment, respectively.

The proposed controller can be re-written as a constrained nonlinear optimization problem, as:
\begin{equation}\label{eq24}
  \begin{array}{c}
  \min\limits_{\Delta M(k+i-1 \mid k)}\quad J=J_{1}+J_{2} \\
  \text {s.t. }\left\{\begin{array}{c}
  \boldsymbol{\chi}(k+i \mid k)=\hat{\boldsymbol{f}}(\boldsymbol{\chi}(k+i-1 \mid k), \Delta M(k+i-1 \mid k)) \\
  \Delta M_{\min } \leq \Delta M(k+i-1 \mid k) \leq \Delta M_{\max }
  \end{array}  \right.\\
  (i=1,2,3)\\
  \end{array}
\end{equation}

In this way, the required external yaw moment for lateral stabilization of the IMDV can be obtained by (\ref{eq24}) using \emph{fmincon} from the \emph{MATLAB} optimization toolbox.

\section{Simulation Validation and Results}\label{sec4}
The RHONN model fidelity is first investigated through simulations under different scenarios. Then, the proposed algorithm is further tested and compared with other methods in two critical situations, for validating its effectiveness. All simulations are conducted on a CarSim/Simulink co-simulation platform, programmed and solved with \emph{MATLAB 2020b}.

\subsection{Simulation validation of the RHONN-based vehicle model}
\begin{figure}[!t]
  \centering
  \includegraphics[width=2.5in]{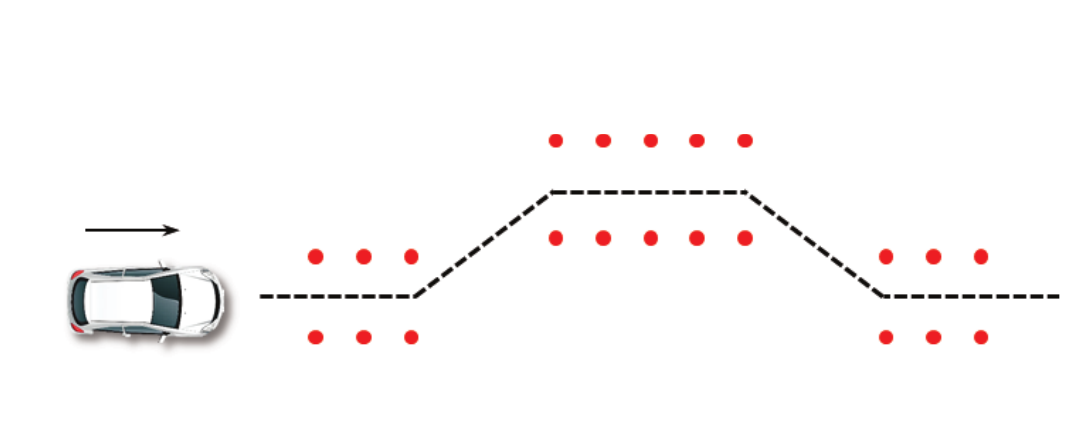}
  \caption{The double-lane change scenario.}
  \label{fig4}
\end{figure}
\begin{figure*}[!t]
  \centering
  \includegraphics[width=6.4in]{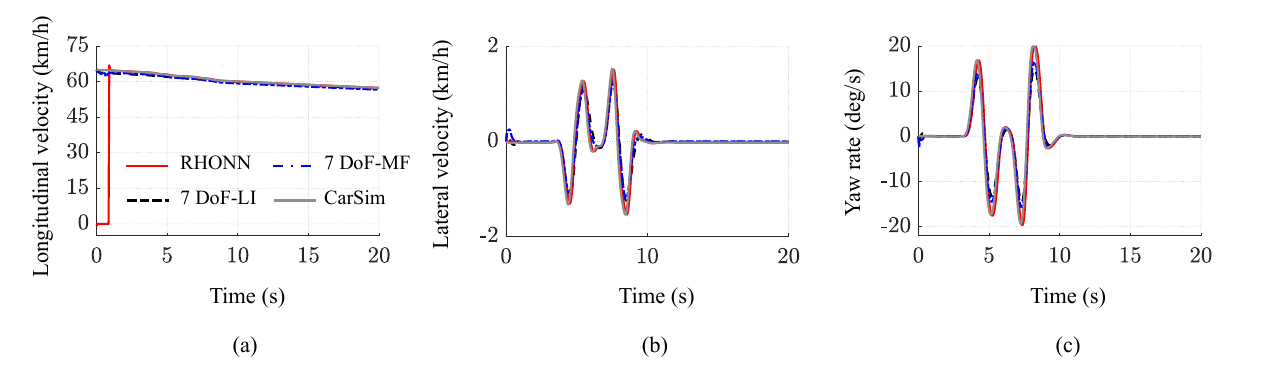}
  \caption{The estimation results under the high-$\mu$ condition. (a)-Longitudinal velocity (b)-Lateral velocity, (c)-Yaw rate.}
  \label{fig5}
\end{figure*}

\begin{figure*}[!t]
  \centering
  \includegraphics[width=6.4in]{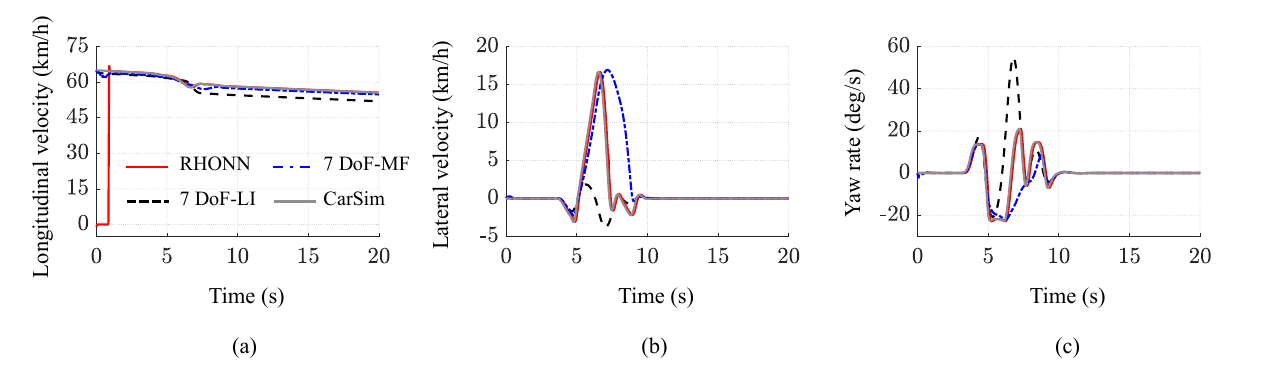}
  \caption{The estimation results under the low-$\mu$ condition. (a)-Longitudinal velocity (b)-Lateral velocity, (c)-Yaw rate.}
  \label{fig6}
\end{figure*}

To validate the fidelity, the state estimation performance of the RHONN model is tested with an IMDV under a double-lane change scenario on CarSim-Simulink co-simulation platform, as shown in Fig. \ref{fig4}. In simulations, the RHONN model, as well as the IMDV physical plant model, receive the same control input signals from the driver model. The actual state vector $\boldsymbol{x}$ is directly from CarSim. Both high-$\mu$ and low-$\mu$ roads are set in simulation, to identify the linear and nonlinear working regions for IMDV, respectively. 

Moreover, the 7 DoF dynamics model is often used to depict a vehicle’s planar motions in a physics-based way, whereas the tyre model is poised as the most challenging aspect. The accuracy of the tyre model determines the physics-based model fidelity. The tyre model can be classified into linear and nonlinear types, tailored to the relationships between tyre motions and forces. The magic formula (MF), a typical nonlinear tyre model, is well-described for tyre dynamics. However, the MF or other nonlinear model has a complex set of empirical parameters and massive calculation steps, which limits its application into practice. To exhibit the metrics of the RHONN-based dynamics model, we conduct the comparisons with the 7 DoF dynamics model, integrating with the MF (7 DoF-MF) and the linear tyre model (7 DoF-LI), respectively. Note further that, in this case, the empirical parameters of the MF and the relevant parameters of the linear tire model are derived by CarSim to ensure the comparability of the simulation results.

The initial longitudinal velocity is set as 65 km/h, without extra driver action on the accelerator or braking pedal. The main parameters applied to the vehicle model are listed in Table \ref{tab1}.

\begin{table}[!t]
  \renewcommand{\arraystretch}{1.4}
  \caption{Vehicle Specifications in CarSim}
  \label{tab1}
  \centering
  \begin{tabular}{lccc}
  \hline
  \bfseries {Body geometry} & \bfseries {Nomenclature} & \bfseries {Value} & \bfseries {Uint}\\
  \hline
  Vehicle mass & $m$ & 2070 & kg \\
  The inertia of the vehicle & $I_z$  & 3658 & kgm$^2$  \\
  Distance from the front axle to c.g. & $l_f$ & 1.362 & m  \\ 
  Distance from the rear axle to c.g.  & $l_r$ & 1.308 & m \\ 
  Track width & $w_B$ & 1.715 & m  \\ 
  Rotational inertia of the wheel & $I_w$ & 2.4 & kgm$^2$  \\ 
  Rolling radius & $r$ & 0.358 & m  \\ 
  Cornering stiffness of the front axle & $C_f$ & -108350 & N/rad \\
  Cornering stiffness of the rear axle & $C_r$ & -105898 & N/rad \\
  \hline
  \end{tabular}
\end{table}

\begin{table}[!t]
  \renewcommand{\arraystretch}{1.4}
  \caption{The statistical errors of the estimations under the high-$\mu$ condition}
  \label{tab2}
  \centering 
  \begin{tabular}{cccc}
  \hline
  \multirow{2}*{} & \bfseries {7 DoF-LI} & \bfseries {7 DoF-MF} & \bfseries {RHONN}\\
  \hline
  ~ & Min / Max / RMSE & Min / Max / RMSE & Min / Max / RMSE \\
  \hline
  $V_x$  & 0.26 / 1.20 / 0.64 &	0.60 / 1.09 / 0.78 & 0.03 / 0.57 / 0.06 \\
  $V_y$  & 0 / 0.77 / 0.20 & 0 / 0.69 / 0.17 &	0 / 0.64 / 0.15 \\
  $\omega_{r}$  & 0.01 / 7.54 / 1.98 & 0 / 6.13 / 1.53 & 0 / 8.07 / 1.96 \\
  \hline
  \end{tabular}
\end{table}

\subsubsection{The high-$\mu$ condition}

The adhesion coefficient for high-$\mu$ road is set at 0.7, indicating that sufficient tyre force can be provided by the ground. This set-up ensures the tires to be work within the linear stable region, which results in the vehicle dynamic behaviour is close to the representation of a linear model. The state estimations are conducted using the proposed RHONN model, the 7 DoF-MF model, and the 7 DoF-LI model. The estimation results further discussed as follows.

According to the results shown in Fig. \ref{fig5}, both the RHONN model and the two 7 DoF models perform well regarding the estimations of the three key state variables. The vehicle starts with 65km/h, while the longitudinal velocity estimation by starts from 0 km/h and soon jumps, converging to the actual value. This transition time of the RHONN model is less than 1 s, which also implies a good adaptation efficiency of the online EKF-based learning algorithm developed in section \ref{sec2.2}.

To further quantify the estimation performance of the different models, statistical errors of the estimations, including the minimum value, maximum value and the root mean square error (RSME), are calculated, respectively.

As Table \ref{tab2} shows, the maximum values of the $V_x$ and $V_y$ estimates under both models are less than 1 km/h, and their RSMEs are lower than 0.8 km/h. The RHONN outperforms both the 7 DoF-MF and the 7 DoF-LI models. It reduces the estimation error of the $V_x$ by 92.3$\%$ and 90.7$\%$, as for the RMSE values compared to 7DoF-MF and 7 DoF-LI, respectively. For the estimates of $\omega_{r}$, the RHONN model and the two 7 DoF models achieve similar performance. Notably, the estimation performance for $\omega_{r}$ of the RHONN model is slightly worse than the two physics-based models, especially for the maximum value. The reason is cast as the time delay for the RHONN model. Concretely, the RHONN model corrects its current outputs by the previous feedback from the real plant, thus introducing delays. The parameters: $\mu_{i}$ and $\beta_{i} (i=1,2, \ldots, n)$ in (\ref{eq4}), are in connection with delayed time. The larger $\mu_{i}$ or $\beta_{i}$, the shorter delays, but it could increase the probability of an overshoot. 

In short, the three models present good state estimation performance on high-$\mu$ condition, exhibiting their abilities on the approximation of vehicle dynamics in linear operation regions. Further, the RHONN model can improve the estimation accuracy, showing its great potential in system dynamics representation.

\subsubsection{The low-$\mu$ condition}
The adhesion coefficient of the low-$\mu$ road is set as 0.35. In this situation, the friction force will not be sufficient, resulting in the tires entering the nonlinear working regions. The state estimation results of the three models under this driving condition are displayed in Fig. \ref{fig7}.

\begin{table}[!t]
  \renewcommand{\arraystretch}{1.4}
  \caption{The statistical errors of the estimations under the low-$\mu$ condition}
  \label{tab3}
  \centering 
  \begin{tabular}{cccc}
  \hline
  \multirow{2}*{} & \bfseries {7 DoF-LI} & \bfseries {7 DoF-MF} & \bfseries {RHONN}\\
  \hline
  ~ & Min / Max / RMSE & Min / Max / RMSE & Min / Max / RMSE \\
  \hline
  $V_x$  & 0.06 / 3.86 / 3.05 & 0.04 / 2.08 / 0.86  & 0.01 / 0.57 / 0.12 \\
  $V_y$  & 0 / 18.62 / 3.97 & 0.01 / 17.92 / 4.14 & 0 / 4.16 / 0.65 \\
  $\omega_{r}$  & 0.01 / 50.27 / 10.22 & 0 / 32.33 / 6.67 & 0 / 11.75 / 2.29 \\
  \hline
  \end{tabular}
\end{table}
According to the results shown in the Fig. \ref{fig7}, in this case, after the initial short-period transition, the output of the RHONN model is more consistent with the actual value provided by CarSim, compared to the results of the two physics-based models. As reflected in Fig. \ref{fig7}(b) and (c), during 5 s to 8 s, there are significant time delay and deviations in the estimation results of the two 7 DoF models. Specifically, the lateral velocity of 7 DoF-MF has almost identical amplitude with the grey baseline but with fewer fluctuations; the 7 DoF-LI shows significant suppression but a larger error on the lateral velocity, whereas its peak yaw rate increases with comparison to the grey baseline. In contrast, the RHONN model achieves a better estimation performance, indicating its strong representation ability of the nonlinear system dynamics. 

The statistical errors of the estimations under the low-$\mu$ condition are listed in Table \ref{tab3}. The maximum errors of the estimations for $V_x $ are 3.86 km/h and 2.08 km/h by 7 DoF-LI and 7 DoF-MF, respectively, while that of the RHONN model are 0.57 km/h.  In terms of $V_y$, the maximum errors of 7 DoF-MF and 7 DoF-LI are 17.92 km/h and 18.62 km/h, but the value for RHONN is 4.16 km/h. The RSME values of the estimations provided by the RHONN model are also lower than that generated by the two 7 DoF models: 0.12 (RHONN) $<$ 0.86 (7 DoF-MF) $<$ 3.05 (7 DoF-LI) km/h for $V_x$ estimation, and 0.65 (RHONN) $<$ 4.14 (7 DoF-MF) $<$ 3.97 (7 DoF-LI) km/h for $V_y$ estimation. Further, regarding the estimation of $\omega_{r}$, the maximum values of the 7 DoF-MF and 7 DoF-LI are 32.33 deg/s and 50.27 deg/s, respectively, compared to the RHONN model at 11.75 deg/s. Similar observation can also be found in the RSME values of the $\omega_{r}$ estimates. These results demonstrate that the RHONN method is advantageous over the two 7 DoF models on state estimation, especially under highly nonlinear operation conditions.

The working regions of the tires determine vehicle dynamics in the planar motion. When the vehicle enters a nonlinear region, the linear tire model can hardly represent the nonlinear relationship between the tire forces and vehicle states. Compared with the linear physics-based model (7 DoF-LI), the nonlinear physics-based model (7 DoF-MF) significantly improves the model accuracy. However, here are two drawbacks: 1) the nonlinear physics-based model can not feature the external disturbances and uncertainties in nature, and the un-modeled dynamics 
exist; 2) the nonlinear physics-based model consists of a plurality of empirical parameters, whereas, it is challenging to acquire the precise value in practical applications.

With these in mind, the metrics for the RHONN-based model are therefore justified. The data-driven RHONN method shows significant superiority in system dynamics modeling by only collecting the inputs and outputs due to its high-order nonlinear combinations of neuron states. The RHONN model achieves real-time identification on the un-modeled dynamics and uncertainties. Moreover, combined with the high-efficiency EKF-based learning algorithm for weight optimization, the proposed RHONN method is proven feasible and effective under linear and nonlinear dynamic conditions.

\subsection{Simulation of the RHONN-based NMPC}
To further test the proposed RHONN-based NMPC for IMDV lateral stabilization, two critical driving scenarios: the double-lane change on a low adhesion road and path tracking on slippery Curvature, are selected for simulation. The driver returns to its original lane after lane change or obstacle avoidance in the double-lane change test. Meanwhile, driving on slippery curvatures, is another dangerous task in real-world applications. Besides, three different ways are outlined herein for comparison:

\begin{enumerate}
\item  {\bfseries{\emph{NMPC-RHONN}}}: The proposed method by this paper. The RHONN model is built to characterize and predict vehicle motions, incorporating with the RHONN-based steady-state responses as intimated above.

\item  {\bfseries{\emph{NMPC-MF}}}: A nonlinear tire model, that is, the Magic Formula (MF), is employed to depict tire dynamics. Thereafter, the 7 DoF dynamics model with the MF is adopted to feature the vehicle behaviors. The linear steady-state responses are taken as the control targets. The nonlinearities of tire forces make the optimization problem (\ref{eq24}) also a nonlinear one, which feasibility has been proven in the existing studies. 

\item  {\bfseries{\emph{LMPC}}}: This is a linear model predictive method. The "linear" means both the steady-state responses and the vehicle dynamics model are linear. The linear vehicle model applies 2 DoF dynamics model embedded with the linear tire model since its simplicity and conciseness for MPC implementation. As such, this method is easy to carry out in real applications.  
\end{enumerate}

\subsubsection{The double-lane change with the low adhesion condition}\label{test1}
\begin{figure}[!t]
  \centering
  \includegraphics[width=3.2in]{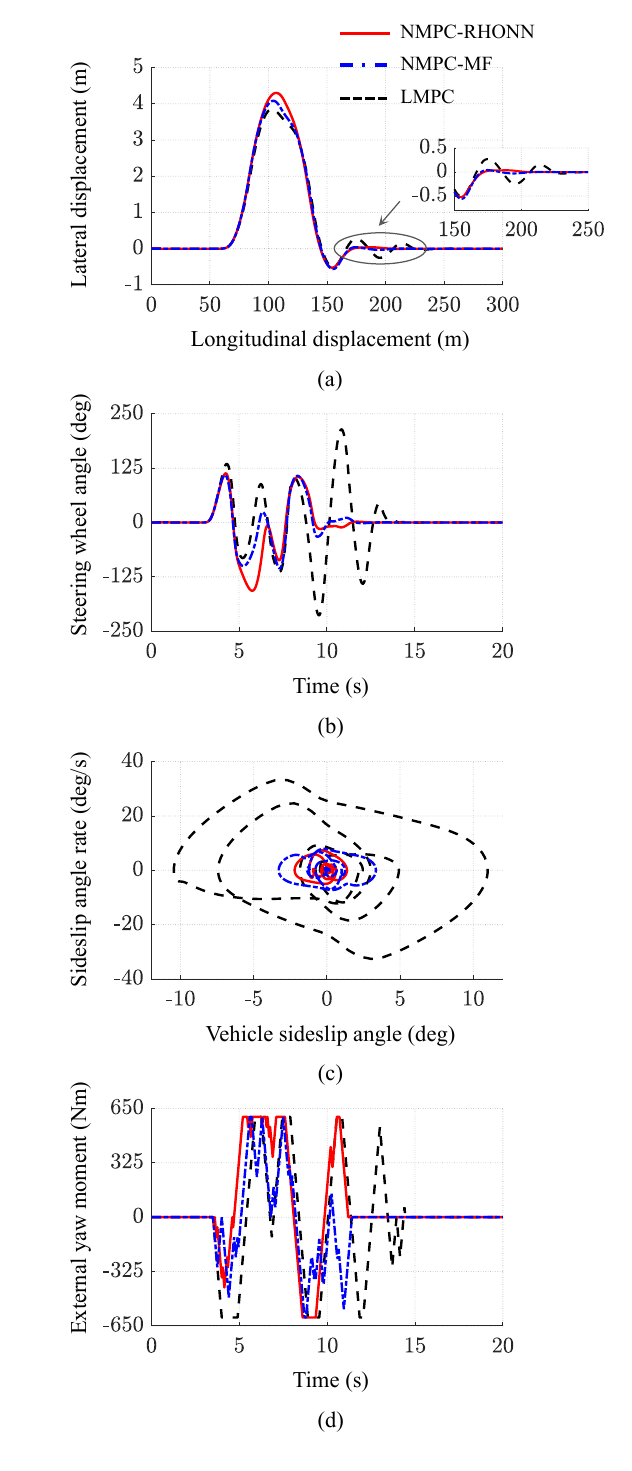}
  \caption{The simulation results of the double-lane change scenario. (a)-Trajectory, (b)-Steering wheel angle,  (c)-Stability phase, (d)-External yaw moment.}
  \label{fig7}
\end{figure}

In this case, the adhesion coefficient of the road is set as 0.35, and the initial longitudinal velocity is set at 65 km/h. No actions are taken from the driver along the testing road. The comparison results of both the proposed and the baseline methods are shown in Fig. \ref{fig7}.

According to the results, under the proposed NMPC-RHONN and the NMPC-MF methods, the IMDV could complete the desired trajectory for double lane change, while the one with LMPC one presents some unstable behaviors. In Fig. \ref{fig7}(b), the LMPC method exerts frequent adjustment on the steering wheel angle and struggles to stabilize the vehicle. The increasing steering wheel angle indicates that the vehicle is experiencing a critical state of instability. In contrast, the NMPC-RHONN and NMPC-MF methods require less control effort and ensure the vehicle under a stable condition. In addition, the sideslip angle - sideslip angle velocity phase $(\beta-\dot{\beta})$ is a necessary way to quantify vehicle’s lateral stability.
The enclosed area under the NMPC-RHONN method is smaller than that of the NMPC-MF and LMPC method in Fig. \ref{fig7}(c), indicating that the vehicle gains a larger lateral stability margin under the proposed approach. Further, LMPC has unexpected oscillations in the external yaw moment in Fig. \ref{fig7}(d). Meanwhile, we test the maximum longitudinal velocity that promises good tracking performance at the starting point of each method. Table \ref{tab4} lists the results:  under NMPC-RHONN, the longitudinal velocity at the starting point of is 74.4 km/h, which improves those under NMPC-MF and LMPC by 2.76$\%$ and 18.47$\%$, respectively. This implies that the NMPC-RHONN method allows higher-speed driving for the IMDV because of its sufficient stability margin. 

\begin{table}[!t]
  \renewcommand{\arraystretch}{1.4}
  \caption{Longitudinal velocity of the IMDV at the starting point} \label{tab4}
  \centering 
  \begin{tabular}{cccc}
  \hline
  \multirow{2}*{} & \bfseries {NMPC-RHONN} & \bfseries {NMPC-MF} & \bfseries {LMPC}\\
  \hline
  Longitudinal Velocity (km/h)  & 74.4 & 72.4 & 62.8 \\
  \hline
  \end{tabular}
\end{table}

\begin{figure}[!t]
  \centering
  \includegraphics[width=3.7in]{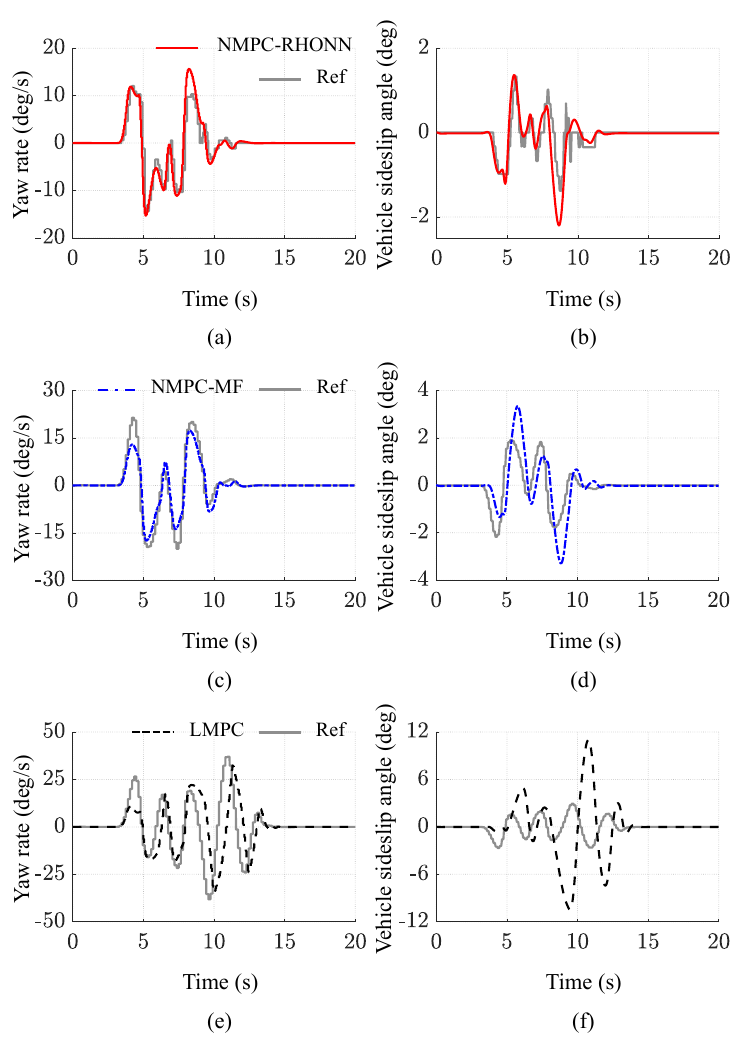}
  \caption{Tracking results of the double-lane change test. (a)- Yaw rate (NMPC-RHONN) (b)- Vehicle sideslip angle (NMPC-RHONN), (c)- Yaw rate (NMPC-MF) (d)- Vehicle sideslip angle (NMPC-MF), (e)- Yaw rate (LMPC) (f)- Vehicle sideslip angle (LMPC).}
  \label{fig8}
\end{figure}

To further illustrate the advantages of the proposed RHONN-based lateral steady-state responses, the results of tracking results are discussed and compared in Fig. \ref{fig8}. The magnitudes and the variations of the desired tracking targets under NMPC-RHONN and NMPC-MF are smaller than those of LMPC, especially for vehicle sideslip angle. This observation is because the steady-state responses of 2 DoF linear model highly rely on the steering wheel angle input. The unexpected sharp changes in the references (Fig. \ref{fig8}(e) and (f)) would be difficult for the tracking, which causes divergence of the systems. Besides, LMPC adopts the linear dynamics model as the predictive model, the model-based vehicle behaviours deviate from the actual dynamics of the IMDV, therefore, resulting in undesired tracking errors. Since the NMPC-MF method employs the nonlinear physics-based model to predict the vehicle states, the overall tracking performance has been improved in Fig. \ref{fig8} (c) and (d). Nevertheless, the linear steady-state responses are taken as the references for NMPC-MF. During 5.3 s to 6.5 s and 8.4 s to 9.7 s in Fig. \ref{fig8}(d), vehicle sideslip angle tracking errors are stark due to the unreasonable targets. In contrast, both the vehicle model's lateral steady-state responses and nonlinear dynamics representation are more accurate by leveraging the developed RHONN. Thus, NMPC-RHONN achieves better tracking performance.
\begin{figure}[!t]
  \centering
  \includegraphics[width=3.2in]{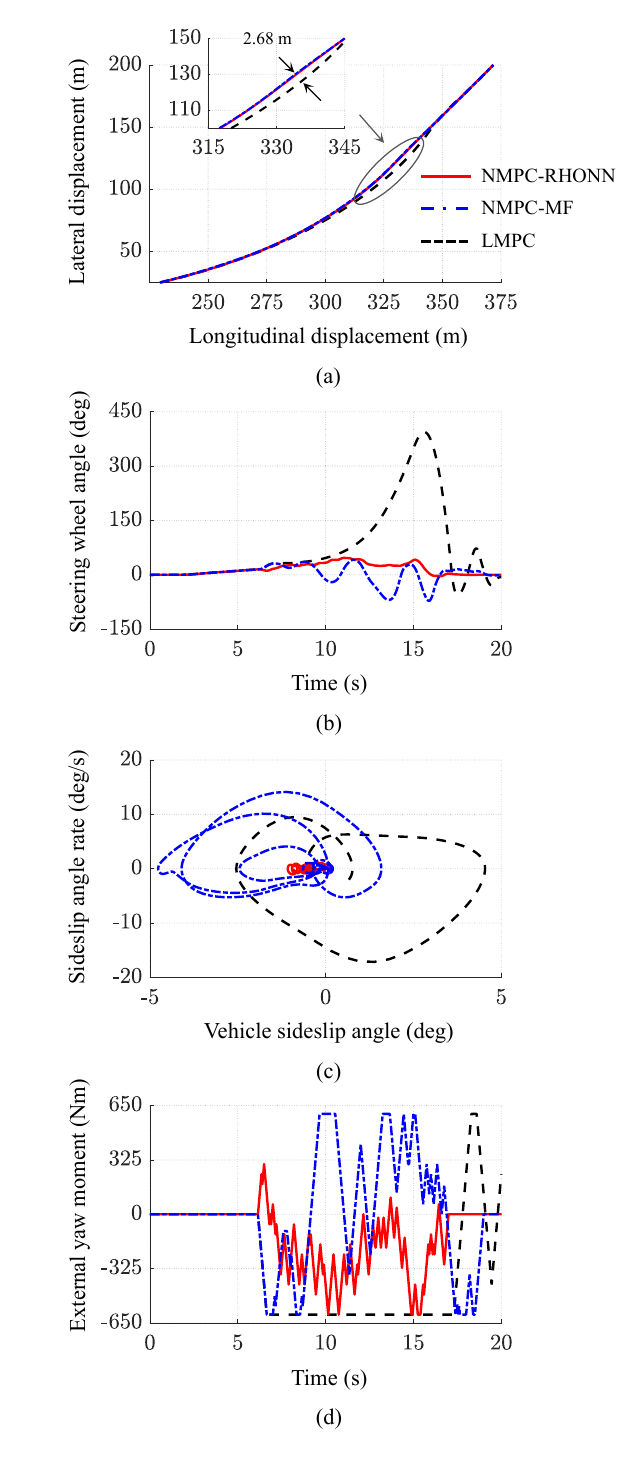}
  \caption{Simulation results of slippery curve test. (a)-Trajectory, (b)-Steering wheel angle, (c)-Stability phase, (d)-External yaw moment.}
  \label{fig9}
\end{figure}

\begin{figure}[!t]
  \centering
  \includegraphics[width=3.7in]{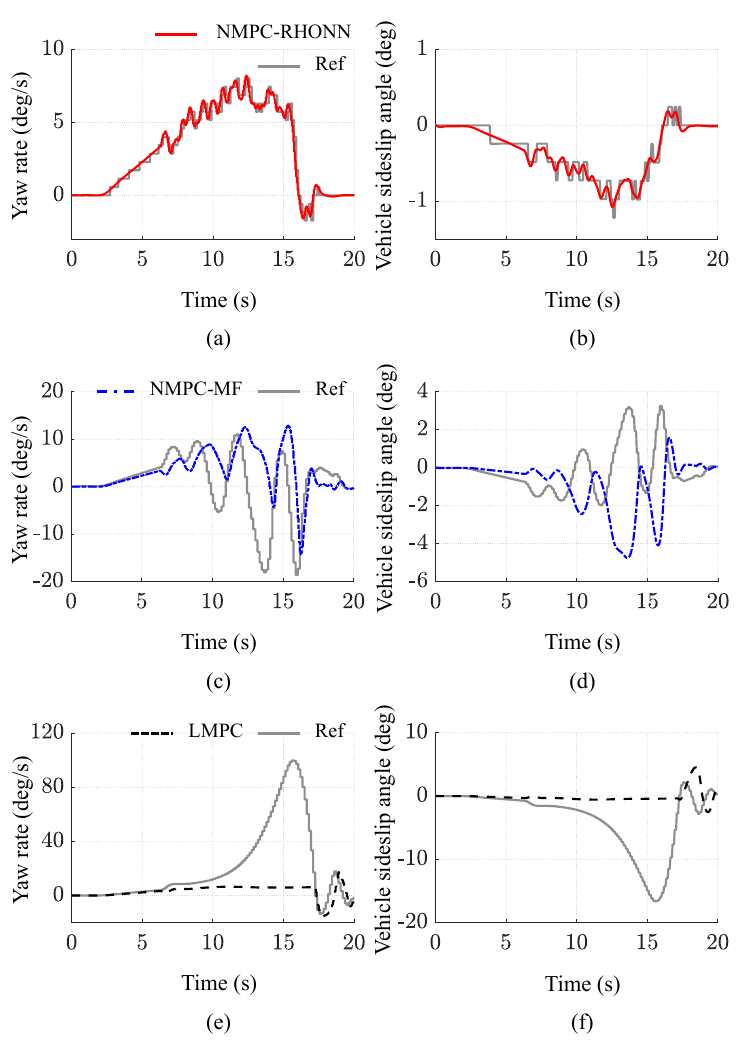}
  \caption{Tracking results of slippery curvature test. (a)- Yaw rate (NMPC-RHONN) (b)- Vehicle sideslip angle (NMPC-RHONN), (c)- Yaw rate (NMPC-MF) (d)- Vehicle sideslip angle (NMPC-MF), (e)- Yaw rate (LMPC) (f)- Vehicle sideslip angle (LMPC).}
  \label{fig10}
\end{figure}

\subsubsection{Path tracking on slippery curves}\label{test2}

In this case, the road adhesion coefficient is set as 0.35 for the slippery curve, and the initial longitudinal velocity of the vehicle is set as 85 km/h. During simulations, the driver model keeps the longitudinal velocity unchanged. Detailed simulation results are reported as follows.

As shown in Fig. \ref{fig9}, during the path tracking task, the lateral stabilization control system is activated after 5 s. Compared to the results of the LMPC method, the vehicle with the two NMPC approaches passes through the slippery curvature under a more stable status; the maximum deviation between the actual trajectory and the reference is negligible, whereas the result is up to 2.68 m with LMPC. The NMPC-RHONN method requires fewer steering wheel angle adjustments than the other two methods. Unlike simulation in \ref{test1}, in this case, the noticeable distinction is made between two NMPC ways. Although the two methods have the same trajectory tracking performance, the enclosed area of NMPC-RHONN is less than NMPC-MF in Fig. \ref{fig9}(c). The phenomenon is still cast as the model fidelity. In fact, the tire is under a combined-slip condition when accelerating and steering simultaneously, different from pure-slip in Section \ref{test1}, and the MF is hard to feature tire dynamics under such condition. Thus, the model fidelity of the nonlinear physics-based model (NMPC-MF) is inferior to the data-driven model (NMPC-RHONN), thus with insufficient lateral stability. The results of Fig. \ref{fig9}(c) illustrate that the NMPC method leads to a smaller $(\beta-\dot{\beta})$ phase area than LMPC, effectively improving the lateral stability.

During the left corning of the IMDV, the applied external yaw moment of NMPC-RHONN is non-positive after 7 s. This mitigates the trend of excessive understeering, caused by the insufficient friction force on the low-$\mu$ road. NMPC-MF has adjusted the lateral dynamics of the IMDV, but both the magnitude and volatility of the external yaw moment are aggressive. By the LMPC strategy, even the steering wheel angle has been increased up to 394 deg at 15.6 s in Fig. \ref{fig9}(b), the unexpected unstable lateral motion cannot be corrected. Besides, the maximum external yaw moment has been conducted since 6.7 s, resulting in the lateral slipping of the vehicle.

The linear steady-state responses of the yaw rate and sideslip angle can hardly be followed using the NMPC-MF and LMPC strategies in Fig. \ref{fig10}. Further, the maximum values of the sideslip angle and yaw rate in Fig. \ref{fig10}(e) and (f) are unacceptable, as they are far beyond the stability bounds. These unexpected results are caused by the discrepancy between the linear model assumption and the nonlinear dynamics of the real system. In contrast, because of the adoption of Algorithm \ref{alg:Solution}, the references are the nearest equilibriums around the previous states, which avoids excessive yaw moment in Fig. \ref{fig9}(d). Consequently, the NMPC-RHONN method has a better tracking performance as before by its high model fidelity and considerable adaption, similar to the previous results in Section \ref{test1}.

\subsubsection{Computational efficiency}
We test the computation burden of three controllers for Section \ref{test1} and section \ref{test2}. Table \ref{tab5} lists the results of two scenarios. The average computational time is 0.0192 ms and 0.0169 ms for NMPC-MF and LMPC, respectively, whereas the proposed NMPC-RHONN controller consumes 0.0169 ms. Here, NMPC-MF takes longer because of its complex physical models, such as the Magic Formula.

The NMPC-RHONN controller has a satisfactory computational efficiency with the network scale designed at a reasonable size. The execution time is less than the controller's sampling frequency of 20 Hz, indicating its real-time applicability. Besides, the computational efficiency can be further improved by optimizing the design of neurons and orders of RHONN.

\begin{table}[!t]
  \renewcommand{\arraystretch}{1.4}
  \caption{The computation time of three controllers on two scenarios} \label{tab5}
  \centering 
  \begin{tabular}{cccc}
  \hline
  \multirow{2}*{} & \bfseries {NMPC-RHONN} & \bfseries {NMPC-MF} & \bfseries {LMPC}\\
  \hline
  Double-lane Change (ms)  & 0.0166 & 0.0193 & 0.0166 \\
  Slippery Curvature  (ms)  & 0.0172 & 0.0194 & 0.0171 \\
  \hline
  \end{tabular}
\end{table}

\section{Conclusion}\label{sec5}

This paper proposes a new nonlinear predictive controller based on the data-driven RHONN modeling method for lateral stabilization of IMDVs. By simulation testing, validation, and comparison analysis, the feasibility and effectiveness of the proposed method are validated. And the following conclusions can be summarized.

\begin{enumerate}
\item The proposed RHONN model, together with the EKF-based learning for weight optimization, can well describe the nonlinear dynamic characteristics of the vehicle. The higher-order polynomial of neuron states are constructed, which features its accurate representation of the nonlinear system dynamics. Hence, the RHONN model achieves a higher fidelity in linear and nonlinear working conditions than the conventional physics-based models.

\item Based on the developed RHONN model, the steady-state yaw rate and sideslip angle are further optimized with partial prior structural knowledge and then set as the desired targets for the low-level tracking controller. In this way, the refined control targets are identified by the "neighbors-searching" algorithm in the feedforward loop, which is close to the previous states. This helps to reduce control efforts and oscillations, avoid system divergence, and enhance the system's tracking control ability and robustness.

\item Further, the NMPC strategy is proposed based on the RHONN model for lateral stabilization of the IMDVs. It improves prediction accuracy and generates a refined external yaw moment for stabilizing the lateral dynamics of the IMDV. The feasibility and effectiveness of the proposed method are conducted in simulations with an IMDV under various critical driving conditions, and the real-time applicability is also justified. Compared with the NMPC-MF and the conventional LMPC methods, the proposed NMPC-RHONN has achieved a better trajectory tracking performance with a more considerable stability margin and a satisfactory computational efficiency, thus improving the lateral stability of the IMDV.
\end{enumerate}


%



\ifCLASSOPTIONcaptionsoff
  \newpage
\fi



%


\bibliographystyle{IEEEtran}
\bibliography{reference} 


\appendix[Entries of High-Order Polynomial Vector]

\begin{small}
$\boldsymbol{\varphi}\left(\boldsymbol{\chi}_{k}, \boldsymbol{u}_{k}\right)=\left[\begin{array}{c}\varphi_{1, k} \\ \varphi_{2, k} \\ \vdots \\ \varphi_{15, k}\end{array}\right]$  

$\boldsymbol{\xi}\left(\boldsymbol{\chi}_{k}, \boldsymbol{u}_{k}\right)=\left[\begin{array}{l}\xi_{1, k} \\ \xi_{2, k} \\ \xi_{3, k} \\ \xi_{4, k}\end{array}\right]=\left[\begin{array}{c}S\left(\chi_{1, k}\right) \\ S\left(\chi_{2, k}\right) \\ S\left(\chi_{3, k}\right) \\ S\left(u_{3, k}\right)\end{array}\right]=\left[\begin{array}{c}S\left(\hat{V}_{x, k}\right) \\ S\left(\hat{V}_{y, k}\right) \\ S\left(\hat{\omega}_{r, k}\right) \\ S\left(\delta_{w, k}\right)\end{array}\right]$

$\left[\begin{array}{l}\varphi_{1, k} \\ \varphi_{2, k} \\ \varphi_{3, k} \\ \varphi_{4, k}\end{array}\right]=\left[\begin{array}{c}\xi_{1, k} \\ \xi_{2, k} \\ \xi_{3, k} \\ \xi_{4, k}\end{array}\right]=\left[\begin{array}{l}S\left(\hat{V}_{x, k}\right) \\ S\left(\hat{V}_{y, k}\right) \\ S\left(\hat{\omega}_{r, k}\right) \\ S\left(\delta_{w, k}\right)\end{array}\right]$

$\left[\begin{array}{c}\varphi_{5, k} \\ \varphi_{6, k} \\ \varphi_{7, k} \\ \varphi_{8, k} \\ \varphi_{9, k} \\ \varphi_{10, k}\end{array}\right]=\left[\begin{array}{l}\xi_{1, k} \xi_{2, k} \\ \xi_{1, k} \xi_{3, k} \\ \xi_{1, k} \xi_{4, k} \\ \xi_{2, k} \xi_{3, k} \\ \xi_{2, k} \xi_{4, k} \\ \xi_{3, k} \xi_{4, k}\end{array}\right]=\left[\begin{array}{l}S\left(\hat{V}_{x, k}\right) S\left(\hat{V}_{y, k}\right) \\ S\left(\hat{V}_{x, k}\right) S\left(\hat{\omega}_{r, k}\right) \\ S\left(\widehat{V}_{x, k}\right) S\left(\delta_{w, k}\right) \\ S\left(\hat{V}_{y, k}\right) S\left(\hat{\omega}_{r, k}\right) \\ S\left(\hat{V}_{y, k}\right) S\left(\delta_{w, k}\right) \\ S\left(\hat{\omega}_{r, k}\right) S\left(\delta_{w, k}\right)\end{array}\right]$

$\left[\begin{array}{l}\varphi_{11, k} \\ \varphi_{12, k} \\ \varphi_{13, k} \\ \varphi_{14, k}\end{array}\right]= \\ \left[\begin{array}{c}\xi_{1, k} \xi_{2, k} \xi_{3, k} \\ \xi_{1, k} \xi_{2, k} \xi_{4, k} \\ \xi_{1, k} \xi_{3, k} \xi_{4, k} \\ \xi_{2, k} \xi_{3, k} \xi_{4, k}\end{array}\right]=\left[\begin{array}{c}S\left(\hat{V}_{x, k}\right) S\left(\hat{V}_{y, k}\right) S\left(\hat{\omega}_{r, k}\right) \\ S\left(\hat{V}_{x, k}\right) S\left(\hat{V}_{y, k}\right) S\left(\delta_{w, k}\right) \\ S\left(\hat{V}_{x, k}\right) S\left(\hat{\omega}_{r, k}\right) S\left(\delta_{w, k}\right) \\ S\left(\hat{V}_{y, k}\right) S\left(\hat{\omega}_{r, k}\right) S\left(\delta_{w, k}\right)\end{array}\right]$

$\varphi_{15, k}=\\\xi_{1, k} \xi_{2, k} \xi_{3, k} \xi_{4, k}=S\left(\hat{V}_{x, k}\right) S\left(\hat{V}_{y, k}\right) S\left(\hat{\omega}_{r, k}\right) S\left(\delta_{w, k}\right)$
\end{small}

%









\end{document}